\documentclass[aps,prb,%
twocolumn,%
groupedaddress,amsmath,amssymb]{revtex4}

\usepackage{amssymb}
\usepackage{amsmath}
\usepackage{graphicx}
\usepackage{subfigure}
\usepackage{textcomp}
\usepackage{color}
\usepackage{amsfonts}
\usepackage{bbold}
\usepackage{dsfont}
\usepackage{epsfig}

\newcommand{\arctanh}[1]{\operatorname{arctan}}

\begin{document}

\title{Controlling the conductance of  molecular wires by defect engineering: {\it a divide et impera} approach} 
\date{\today} 
\author{Daijiro Nozaki$,^1$ Horacio M. Pastawski,$^2$
and Gianaurelio Cuniberti$^1$}
\email{g.cuniberti@tu-dresden.de}
\affiliation{$^1$Institute for Materials Science and Max Bergmann Center of Biomaterials, 
Dresden University of Technology, D-01069 Dresden, Germany\\
$^2$Instituto de F\'{i}sica Enrique Gaviola (CONICET-UNC) and Facultad de Matem\'{a}tica, Astronom\'{i}a y  F\'{i}sica, Universidad Nacional de C\'{o}rdoba, 5000 C\'{o}rdoba, Argentina
}
\begin{abstract}
Understanding of charge transport mechanisms in nanoscale structures is essential for the development of molecular electronic devices.
Charge transport through 1D molecular systems connected between two contacts is influenced by several  parameters such as the electronic structure of the molecule and the presence of disorder and defects.  In this work, we have modeled 1D molecular wires connected between electrodes and systematically investigated the influence of both soliton formation and the presence of defects on properties such as the conductance and the density of states. Our numerical calculations have shown that the transport properties are highly sensitive to  the position of both solitons and defects. 
Interestingly, the introduction of a single defect in the molecular wire  which divides it into two fragments  both consisting of an odd number of sites creates a new conduction channel in the center of the band gap resulting in higher zero-bias conductance  than for defect free systems. This phenomenon suggests  alternative routes toward engineering molecular wires with enhanced conductance.
\end{abstract}


\maketitle

\section{Introduction}
The investigation of quantum transport through molecular systems has become an important research field in the last few decades.\cite{1Gio-Book05}  Progress in measuring and fabrication techniques\cite{Agrait} has led to the continuous miniaturization of electronic devices, which have reached the point where quantum effects are important.  For example,  semiconductor devices have been reduced in size to the nanoscale\cite{Hisamoto00,Pai-Chun06} and even to the atomic scale.\cite{Huang02,Singh06}  In addition to the report of metal atomic wires,\cite{Takayanagi,Yanson,Smit,Bettini,Ho-Chain}  stable and rigid carbon atomic chains have been reported recently.\cite{Cchain,Chuvlin} 
A key idea behind the advances in the understanding of charge transport through molecular systems is based on the view proposed by Landauer, \textit{conductance is transmission}.\cite{Landauer,Buttiker}    

Potential applications of molecular devices range from novel computer architectures\cite{Ellenbogen,Joachim} to chemical sensors\cite{Cui01} and medical diagnostics \cite{Patolsky04}.
Among the various molecular devices, 1D conductors such as molecular wires,\cite{Moleuclar-wire,Israel-STM,NDLang}  have been considered to be one of the most fundamental components for nanotechnology. Due to the reduction of size and dimensions of materials, 1D systems show sensitive response to external field or intrisic characteristics, which can be exploited for the development of molecular devices such as biological sensors\cite{Cui01}   with high sensitivity. 

Polymers are one of the most promising materials for acting as 1D conductors. Their applicability\cite{Ubiquitous}  ranges from  displays\cite{PPV,OLED,Jenekhe} to thin film transistors,\cite{TFT}  photovoltaics,\cite{Graetzel}  and solar cells.\cite{Gunes,Blom}   The compatibility of polymer materials with light-weight, mechanically flexible plastic substrates and  new fabrication methods make them possible candidates for future electronic devices.  

Intrinsic properties of molecules as well as external perturbations can strongly affect the transport properties of the resulting low dimensional devices.  For example,  the conductance of  1D polymers drastically depends on  the concentration and  position of impurities (dopant)\cite{Chiang77,Chiang78}  or defects.\cite{Brian}   Other effects which are important include dimerization and the formation of solitonic defects.    Hence, all of  these properties  should be considered in the design of 1D devices. 
 
Therefore, in order to facilitate the design of controllable molecular devices, it is necessary to use a theoretical approach that simplifies the monitoring of how these intrinsic features affect their transport  properties. Such approach should allow one to gain understanding and deep insights into the  physical origins of the behavior of these materials. The knowledge acquired through these models can be used for the interpretation and elucidation of experimental observations, as guidelines for the planning of 
experiments, as well as in the design of molecular devices.

In this paper we model molecular junctions where 1D molecular wire is connected between two electrodes using a tight-binding approach. Then we  systematically investigate how the degree of dimerization, and the position of solitonic and binding defects affect the transport properties of such 1D  systems. We  calculate the electronic density of states (DOS) and the conductance using the Landauer model in terms of  the equilibrium Green functions.

This article is divided into the following sections. 
Section II briefly presents the theoretical framework and
 section III presents  the effect of solitonic and binding defects on quantum transport through molecular systems, respectively. 
 Finally section IV summarizes this article, stressing on the importance of the defect position on the linear conductor.

The effect of an odd vs.~an even number of sites in the DOS and in the conductance is summarized in Supplementary Information A, and the effect of dimerization is summarized in Supplementary Information B. In addition, the relationship between dimerization and length dependence of conductance is summarized in Supplementary Information C. 

\section{Theoretical framework}
\begin{figure}[ht!]
\begin{center}
\includegraphics[width=8.5cm,clip=true]{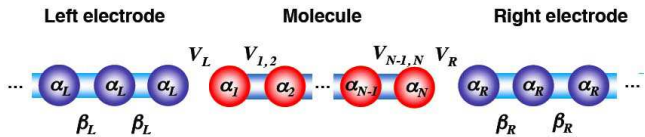}
\end{center}
\caption{\small{ Tight-binding description of the molecular junction considered in this work. The on-site energy terms, $\alpha_{n}; n=1, 2, \cdots, N, {\rm L}, {\rm R}$, are set to 0  for the simplicity. The $\beta_{{\rm L/R}}$ are hopping integrals for left/right 1D electrode. $V_{{\rm L/R}}$ and $V_{n, n+1}; n=1,2, \cdots, N-1$ are transfer integrals for the molecule/electrode interface in the left/right side and hopping integral between nearest neighbors, respectively. The detail of theoretical framework is shown in Appendix A.}}
\label{Fig1}
\end{figure}

 We model a molecular junction by connecting a 1D molecular wire between two electrodes.  
 We describe this molecular system using a standard tight-binding Hamiltonian\cite{SSH79,Mujica1,Mujica2,PAW95}  $H_{\rm C}$ that considers only $\pi$ orbitals. Here, the dimerization is represented by a sequence of alternating weak and strong bonds while the solitonic defect is described by  a pair consecutive weak bonds. For clarity, we model the electrodes as 1D wires with Hamiltonians $H_{\rm L}$ and $H_{\rm R}$, which result in a fair representation of infinite reservoirs. Fig.~\ref{Fig1} shows the molecular junction considered in this work. The tunneling between the central molecule and
the left and right electrodes is represented by two matrix elements $V_{{\rm L}}$ and $V_{{\rm R}}$.
Note that the electrode parameters and tunneling amplitude reduced to  into two independent parameters $\Gamma_{\rm L}$ and $\Gamma_{\rm R}$; the escape rates to the left and to the right
electrode, respectively.\cite{PM01} The details this of theoretical framework are described in Appendix A.

In principle, we can expand this method to general molecular systems which include any type of orbitals. In this work, we simplified the model Hamiltonian by only considering $\pi$-orbitals since the contribution from other orbitals to the conductance near the Fermi energy is negligible in the sp2 carbon systems which we addressed. However, in the case of the molecular wires including metal atoms,\cite{Tyler,Bera} where s-, p-, and/or d-orbitals play a role in charge transport, those orbitals have to be taken into account.
 

 \section{Result and discussion}

\subsection{Influence of the position of soliton}
Dimerization and soliton formation are closely related processes which have been extensively studied. The dimerization has been examined the first in the context of metal-insulator transition where dimerization is known as Peierls distortion,\cite{Peierls} while the soliton formation has been examined in the field of conducting 1D polymers.\cite{Conwell} These effects  strongly modify the conductance and other electronic properties of  1D polymers. Thus, such effects in single molecular wires connected between contacts also should be considered.
In order to estimate the influence of solitons on transport, we modeled a 1D molecular system consisting of an odd number of sites coupled with electrodes, and examined the dependence of the transmission spectra, total DOS (TDOS) and local DOS (LDOS) on the position of a soliton. For  simplicity we assumed that there is a single soliton on a molecular wire. However, this approach can be also applied to larger numbers of solitons  on molecular wires.  
\begin{figure}[ht!]
\begin{center}
\includegraphics[width=8.5cm,clip=true]{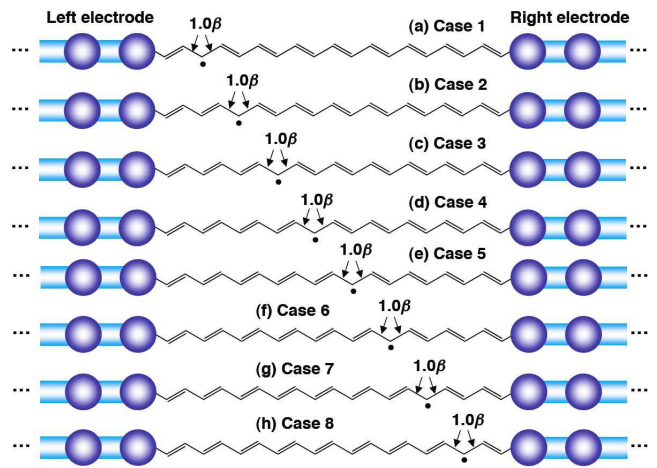}
\end{center}
\caption{\small{ Schematic description of  poly-acetylene-based (PA-based) molecular wires consisting of 19 sites, with single solitonic defect, connected between two 1D electrodes.  Depending on the position of solitonic defect, 8 cases can be considered. Transfer  integrals for double bonds and single bonds are set to $V_{\rm d}=1.2\beta$ and $V_{\rm s}=0.8\beta$, respectively. Two hopping parameters nearby the solitonic defect are set to $1.0\beta$. Coupling constants are set as $V_{{\rm L/R}} =0.8\beta$. Left/right 1D electrodes are treated by Newns-Anderson model\cite{Anderson1,Anderson2}  with $\alpha_{{\rm L/R}}=0$ and $\beta_{{\rm L/R}}=\beta$.    }}
\label{Case18}
\end{figure}
\begin{figure*}[ht!]
\begin{center}
\includegraphics[width=15cm,clip=true]{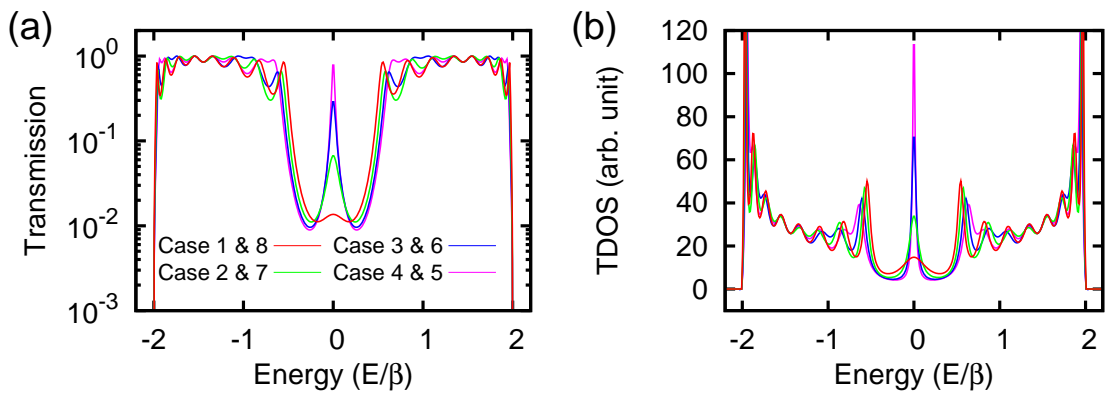}
\end{center}
\caption{\small{  (a) Transmission spectra and  (b) TDOS of the molecular system consisting of 19 sites  connected between two leads parameterized by the positions of a soliton.  }}
\label{Soliton-TMDOS}
\end{figure*}
%
\begin{figure}[ht!]
\begin{center}
\includegraphics[width=8.5cm,clip=true]{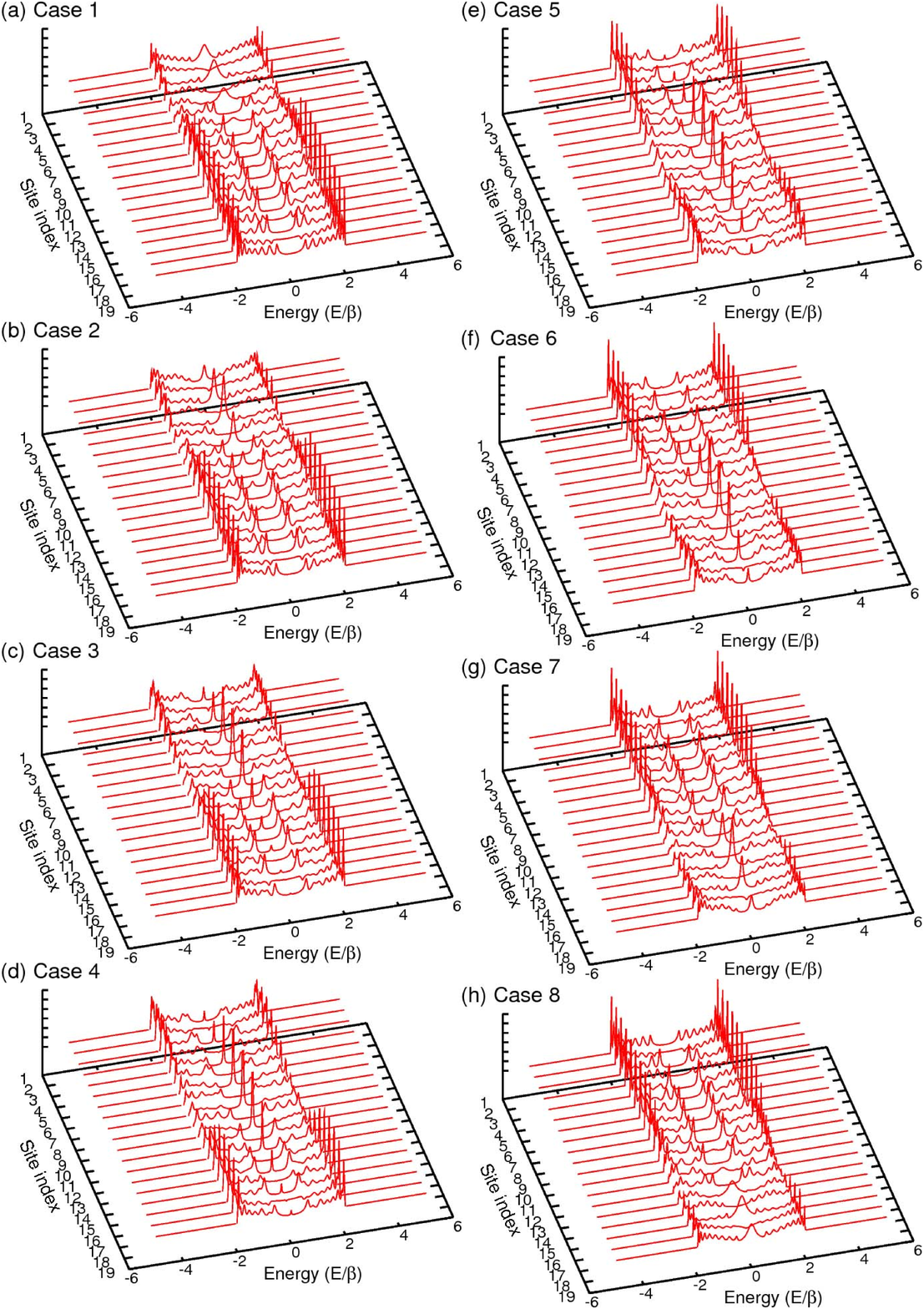}
\end{center}
\caption{\small{ Surface plot of LDOS of  molecular wire shown in Fig.~\ref{Case18} as a function of energy and site index.  The localized peaks of LDOS at $E=0$ migrate along the wire in conjunction with the position of soliton in Fig.~\ref{Case18}.    }}
\label{Soliton-walk}
\end{figure}

 Figure \ref{Case18} shows modeled poly-acetylene-based (PA-based) molecular wires consisting of 19 sites connected between 1D electrodes and allowing for possible 8 soliton positions along the molecular wires.  Transfer  integrals associated with double bonds and single bonds are set to $V_{\rm d}=1.2\beta$ and $V_{\rm s}=0.8\beta$, respectively. The  two hopping parameters on either side of the soliton site are set to $1.0\beta$. All on-site energies are set to $\alpha_n = 0; n=1,2, \cdots, N$. The coupling constants to the electrodes are set as $V_{{\rm L/R}} =0.8\beta$.  The  1D electrodes are treated by the Newns-Anderson model\cite{Anderson1,Anderson2} with $\alpha_{{\rm L/R}}= 0$ and  $\beta_{{\rm L/R}}=\beta$. Fermi energy is set to $E_{\rm F}= 0$. 

Fig.~\ref{Soliton-TMDOS} shows calculated transmission spectra and TDOS of the 1D molecular systems  with different soliton positions. 
Interestingly, depending on the position of the soliton, the transmission at  the Fermi energy ($E=E_{\rm F}$) changes drastically (See animation in Supplemental Information).  When the soliton lies near an electrode, the resonances in the transmission spectra and TDOS at $E=E_{\rm F}$  broaden with lower peaks.   In this case, in spite of having an odd number of sites, the features of the TDOS and transmission resemble those for the  PA-based molecular junctions 
 which have  an even number of sites (See Fig.~S4(a) or (c) in Supplemental Information).  On the other hand, when a soliton lies in the middle of the molecular wire, the resonances in the transmission spectra and TDOS sharpen and get narrower.  The value of the transmission at $E=E_{\rm F}$ approaches its theoretical maximum of 1.0.

In order to elucidate the relationship between the positions of a soliton and sharpness of resonant peaks, the distribution of LDOS along the molecular framework was  calculated. 
Figure \ref{Soliton-walk} shows the LDOS surface plot as a function of energy and site index. The LDOS surface plot has peaks near the Fermi energy ($E=E_{\rm F}$) where the soliton is located.
The LDOS peaks at $E=E_{\rm F}$ broaden and reduce in height when the soliton is located at the terminal of the molecular wire  (Fig.~\ref{Case18}(a) and (h)), whereas the LDOS peaks at $E=E_{\rm F}$ sharpen and narrow when the soliton is located in the middle of molecule (Fig.~\ref{Case18}(d) and (e)).  

This feature can be explained as follows. 
When a soliton lies at the terminal of a molecular wire, the localized state on the molecular wire strongly interacts with surface states of the lead broadening the LDOS peaks around $E=E_{\rm F}$. This means an electron occupying the soliton state easily escapes back to the electrode where the electron was injected from, so that the chance of the electron traveling through to the other electrode is reduced resulting in low transmission at $E=E_{\rm F}$. Meanwhile, when a soliton lies at the center of a molecular wire, the localized state of the molecular wires interacts symmetrically and weakly with surface states of the two leads producing sharp LDOS peaks.

 This conclusion can also be understood in terms of  the generalized symmetry condition.\cite{Azbel,Luis,DPW89,LPD90}  The transmission peak close to the resonance is approximately described as 
$T_{{\rm max}}(E\sim\tilde{E_0})=4\Gamma_{{\rm L}}\Gamma_{{\rm R}}/(\Gamma_L+\Gamma_R)^2$, where $\tilde{E_0}$ is the energy of the molecular orbital associated to the soliton. $\Gamma_{{\rm L/R}}$  follows a simple relation, $\Gamma_{{\rm L/R}}\propto |V_{{\rm L/R}}| |\Psi_{{\rm L/R}}|^2$, where $\Psi_{{\rm L/R}}$ is the orbital amplitude of the molecule at the left/right interface.  Thus, symmetric coupling of the molecule with two electrodes, $\Gamma_{{\rm L}}=\Gamma_{{\rm R}}$, which corresponds to the soliton localization at the center of the molecule in case 4 or 5 in Fig.~\ref{Case18}, gives optimal transmission probabilities, while asymmetric coupling, as can be seen in case 1 or 8 in Fig.~\ref{Case18}, gives lower transmission probabilities.

\subsection{Creation of transmission channels by a defect}
We have examined how the position of a soliton modifies the conductance of molecular wires and seen that the conductance is greatly reduced when the soliton lies near electrodes. Here, we investigate the influence of the defect and  demonstrate the counter-intuitive result that the introduction of a defect satisfying special conditions creates a transmission channel leading to higher conductance than in defect free systems.

In general, realistic molecular junctions will contain defects. The defects  may induce disorder in the molecular framework. In the presence of defects both  the on-site energies and the hopping integrals are modified. Therefore, the transport properties of the molecular junctions will also be changed. Thus we need to consider the effect of defects on quantum transport.  
We will now examine the dependence of the transmission and DOS on the position of the defect. Here we only consider fluctuations in the hopping integrals and introduce a defect as a reduced transfer integral in a molecular system with an even number of sites. For simplicity we consider a single defect in the molecular system. 
\begin{figure}[ht!]
\begin{center}
\includegraphics[width=8.5cm,clip=true]{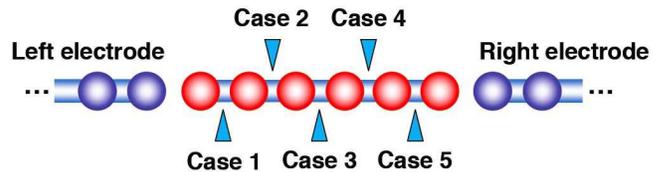}
\end{center}
\caption{\small{ Schematic description of a non-dimerized 1D linear molecule consisting of 6 sites, with a single binding defect, connected between two electrodes.  Depending on the positions of the binding defect, 5 cases can be considered (Case 1 - 5).  Transfer integrals for nearest neighbors and coupling strength at left/right interface are set to $V_{n,n+1}=1.0\beta; n=1,2,\cdots, N-1$ and $V_{{\rm L/R}} =\beta/2$, respectively. The transfer integral corresponding to the position of the binding defect is reduced to $\beta/2$.     }}
\label{defect-case}
\end{figure}

\begin{figure*}[ht!]
\begin{center}
\includegraphics[width=15cm,clip=true]{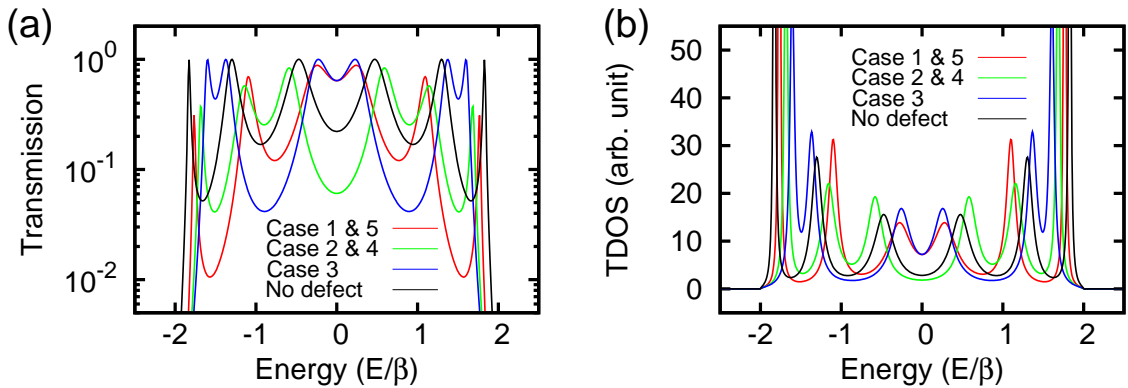}
\end{center}
\caption{\small{ (a) Transmission spectra and  (b) TDOS of the molecular system consisting of 6 sites parameterized by the position of a defect.  }}
\label{defect-TMDOS}
\end{figure*}

Figure \ref{defect-case} shows the schematic description of 1D molecular wire consisting of 6 sites, with a single defect, coupled  to two  electrodes. Transfer integrals for nearest neighbors and coupling strength at the  left and right interfaces are set to $V_{n,n+1}=\beta$ and $V_{{\rm L/R}}=\beta/2$, respectively.  The defect is introduced as a reduction of the transfer integral $V$ between the two sites on either side of the defect to a value of $\beta/2$. We considered 5 cases of  different positions  of a single defect on a molecular wire and investigated position-dependence of the transmission and DOS of the molecular wire. 
\begin{figure}[ht!]
\begin{center}
\includegraphics[width=8.5cm,clip=true]{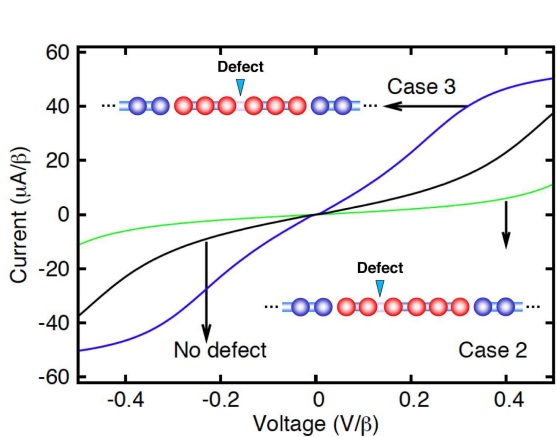}
\end{center}
\caption{\small{ I-V curves of the molecular wires consisting of 6 sites with a single defect. The I-V curves of the molecular wires with a single defect for case 2 and 3 are shown in green and blue, respectively. As a reference, I-V curve for the defect free molecular wire is shown in black.  }}
\label{IV}
\end{figure}
\begin{figure}[ht!]
\begin{center}
\includegraphics[width=8.5cm,clip=true]{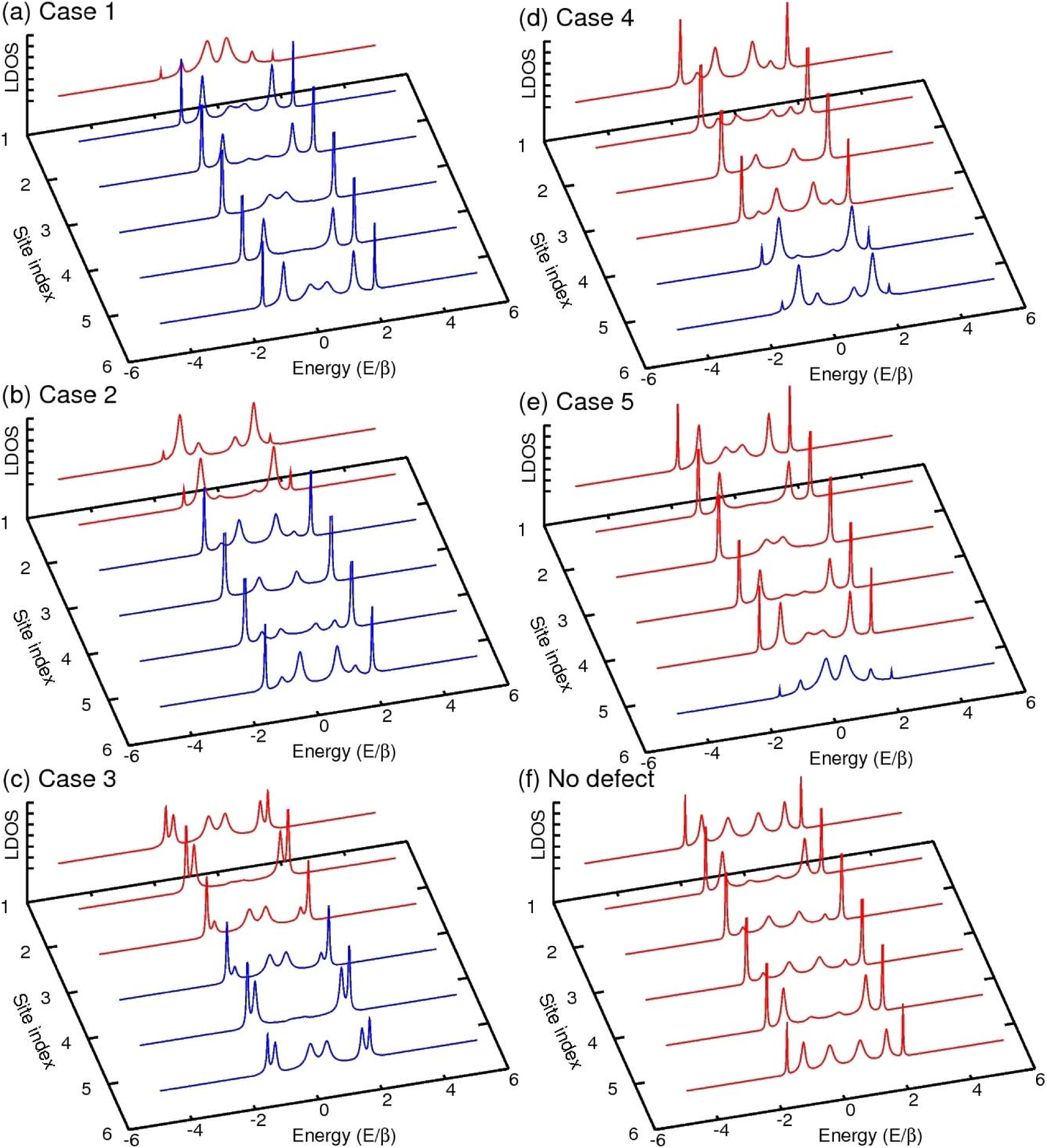}
\end{center}
\caption{\small{ Surface plot of LDOS of  molecular wires with a single defect as shown in Fig.~\ref{defect-case} as a function of energy and site index.  In each case LDOS plots are depicted in different colors across a defect.    }}      
\label{defect-LDOS}
\end{figure}   

Figure \ref{defect-TMDOS} shows the transmission spectra and TDOS of the molecular wire shown in Fig.~\ref{defect-case}. Depending on the position of the  defect, the transport properties change dramatically particularly around $E=E_{\rm F}$. Fig.~\ref{IV} shows I-V curves of the molecular wire calculated in the linear response regime. 
Surprisingly, the transmission at $E=E_{\rm F}$ in cases 1, 3, and 5 is higher  than for defect free systems (Fig.~\ref{defect-TMDOS}). 

In order to further analyze this behaviour, we calculated the LDOS as a function of energy and site index. 
Figure \ref{defect-LDOS} shows the surface plot of LDOS as a function of electronic energy and site index. The position of the defect is illustrated by changing the colour of the LDOS curve from red to blue on either side.   When a defect is introduced such that the molecular wire is divided into two fragments which have an odd number of  sites, for example $3+3$, the LDOS plot of each fragment, as shown  in Fig.~\ref{defect-LDOS}(c), resembles that of a linear chain having 3 sites ($N=3$) without defects, as shown  in Fig.~3(a). The LDOS in Fig.~\ref{defect-LDOS}(c)  shows non-negligible split peaks around $E=E_{\rm F}$. Likewise when a defect is introduced such that  the molecular wire is divided into fragments containing $1+5$ or $5+1$ sites, the LDOS plot of longer fragment resembles that of a linear chain having 5 sites ($N=5$) without defects as can be seen in Fig.~3(b). The LDOS in Fig.~\ref{defect-LDOS}(a) and (e)  show non-negligible split peaks around $E=E_{\rm F}$. This is why, despite including the single defect, these cases lead to higher transmission probabilities and TDOS around $E=E_{\rm F}$  than the defect free systems shown in Fig.~\ref{defect-TMDOS} and Fig.~\ref{IV}.  

On the other hand, when the defect is introduced such that the molecular wire is divided into two fragments both having an even number of  sites, for example $2+4$, the LDOS plot of each fragment resembles that of a linear chain having an even number of  sites without defects. The LDOS in Fig.~\ref{defect-LDOS}(b) and (d) are small around $E=E_{\rm F}$. Thus, these cases lead to low transmission probabilities and TDOS around $E=E_{\rm F}$ as shown in Fig.~\ref{defect-TMDOS}.

\section{Conclusion}
In summary, in order to elucidate the factors controlling charge transport through 1D molecules connected between electrodes we have modeled molecular junctions using the tight-binding method. Then we calculated the the influence of the parameters including molecular lengths, degree of dimerization, odd-even effects,  soliton formation, and defects on transport properties of the molecular junctions at equilibrium using the Landauer approach combined with the Green's function formalism. 

The numerical calculations have shown that the transport properties at the Fermi energy ($E_{\rm F}=0$) dramatically change depending on the degree of dimerization and on whether the number of the sites is odd or even.   It has been shown that dimerization of the molecular wires associated with Peierls distortion reduces the DOS at the Fermi energy, leading to exponential decay of length dependence of conductance of the molecular wires. We also proved that the damping factor is closely related to  the degree of dimerization of the molecular wires.  Interestingly, the longer chain system without dimerization showed no decay in the conductance at the Fermi energy. In the case of molecular wires with  an odd number of sites, the extra non-bonding state appears in the middle of the band gap contributing to high conductance at low bias. Additionally, in the molecular wires with an odd number of sites, the conductance at the Fermi energy was independent of the length of the molecular wires. 

Our calculations  also demonstrate that the transport properties are highly sensitive to the position of solitons and defects. When a soliton lies near one electrode it strongly interacts with surface states of that electrode giving rise to a low, broad peaks in the zero-bias conductance, while when a soliton lies in the middle of the molecular wire, the interaction of the localized state with the electrodes is weaker  leading to high conductance at the Fermi energy with sharp resonant peak. Concerning on the calculation of the influence of defects, we obtained a counter-intuitive result that in some cases defects increase the zero-bias conductance. Our findings are that a defect which divides the molecular wire into two fragments both having an odd number of  sites creates a new conduction channel and enhances the zero-bias conductance. Since thermal fluctuation naturally creates this sort of defects that, at least for a while, strongly favors charge transfer, we might infer that this is one of the underlying mechanisms of transport at room temperature.  We believe that the presented systematic study could be predictive for a wider material ({\it e.g.} semiconductor/organic interfaces). Our results could guide the synthesis of novel molecular wires with enhanced conductors.

\section*{Acknowledgments}
This work was partially funded by the Volkswagen Foundation and by the WCU (World  Class University) program through the Korea Science and Engineering Foundation funded by the Ministry of Education, Science and Technology (Project No. R31-2008-000-10100-0), and the European Social Funds in Saxony 
and the cluster of excellence "ECEMP -- European Centre for Emerging 
Materials and Processes Dresden" within the excellence initiative of the 
Free State of Saxony. We acknowledge the Center for Information Services and High Performance Computing (ZIH) at the Dresden University of Technology for  computational resources.  The work of Horacio M. Pastawski was possible under the Eramus Mundus  Master's Program at Biotechnologische Zentrum (BIOTEC). We thank Claudia Gomes da Rocha, Cormac Toher, and Luis Fo\`{a} Torres for useful discussions.

\appendix

\section{Theoretical framework}

We model a molecular junction by connecting a 1D molecular wire  between two electrodes. For clarity  we model  the electrodes as 1D wires. We describe this system using a tight-binding Hamiltonian, considering only $\pi$ orbitals. The tight-binding parameters used depend explicitly  on the interatomic distances in real space. Thus, we can introduce defects or disorder by directly controlling on-site energies (diagonal elements) and hopping integrals (off-diagonal elements) of the electronic tight-binding Hamiltonian matrices.  

Figure 1 shows the molecular junction considered in this work. Our starting point is to divide the entire 1D system into three regions; left electrode, right electrode, and central molecular region. We write the electronic Hamiltonian as
\begin{equation*}   
    H = H_{\rm L}+V_{{\rm L}}+H_{\rm C}+V_{{\rm R}} + H_{\rm R},
\end{equation*}
where $H_{{\rm L/R}}$, $H_{\rm C}$, and $V_{{\rm L/R}}$ are respectively the Hamiltonian matrices for the left/right electrodes, for the the central molecular region, and the matrix representing the coupling between the central molecule and the left/right electrodes. In this case, we considered only nearest neighbor interactions, so that the terms $V_{{\rm L/R}}$ are scalar values. The matrix elements of $H_{\rm C}$ are given by  $[H_{\rm C}]_{m,n}=\alpha_{n}\delta_{m,n}+V_{m,n}(1-\delta_{m,n})$, where $\alpha_n$ and $V_{m,n}$ are on-site energies and hopping integral between $m$-th and $n$-th atomic orbitals, respectively. In this study, the electrodes are approximated as semi-infinite 1D chains with a single-orbital per site, thus the Hamiltonians of the semi-infinite electrodes $H_{{\rm L/R}}$ have similar forms with $H_{\rm C}$. For simplicity we only consider the interaction between nearest  neighbors.  The matrix elements of left/right electrodes are given by $[H_{\rm L/R}]_{m,n}=\alpha_{{\rm L/R}}\delta_{m,n}+\beta_{L/R}(1-\delta_{m,n})$.
Hereafter we normalize all hopping integrals, $\beta_{{\rm L/R}}$ and $V_{m,n}$, in terms of $\beta$ and  set all on-site energies as 0, $\alpha_n=0; n=1, 2, \cdots, N, {\rm L}, {\rm R}$. 

In the orthogonal basis the retarded Green function is given by 
\begin{equation*}   
    G^{\rm{R}}(E) = [(E+ {\rm{i}} \eta)I-H_{\rm{C}}  -\Sigma_{\rm{L}}(E)- \Sigma_{\rm{R}}(E) ]^{-1},
\end{equation*}
where $I$, $\eta$, and $\Sigma_{{\rm {L/R}}}(E)$ are the identity matrix, the positive infinitesimal, and the self energy in the left/right electrode, respectively. The self-energy term  $\Sigma_{{\rm L/R}}(E)$  is defined as $\Sigma_{{\rm L/R}}(E)=V_{{\rm L/R}}g_{{\rm L/R}}(E)V_{{\rm L/R}}^\dag$,
  where $g_{{\rm L/R}}(E)$ is the surface Green function of left/right 1D electrode. The left/right electrode consists of  equally spaced 1D sites and coupling between nearest neighbors is set as $\beta_{{\rm L/R}}=\beta$.  The surface Green function of the electrode consisting of  equally spaced 1D sites considering only nearest neighbor interaction can be analytically described by the Newns-Anderson model\cite{Anderson1,Anderson2}  as $g_{{\rm L/R}}(E)=\exp{( {\rm{i}} k)/\beta_{{\rm L/R}}}$ whose derivation\cite{Emberly} is summarized in Appendix B.

The conductance of a molecular junction in a low bias is estimated from Landauer's formula $G=G_0T(E)$, where $G_0=2\textrm{(for spin)}\frac{e^2}{h}$ is the conductance quantum and $T(E)$ is  the electronic transmission probability. $T(E)$ is obtained from the Fisher-Lee relation\cite{Fisher-Lee}: 
$T(E) = \textrm{Tr}[ G^{\rm R}(E) \Gamma_{\rm L}(E) G^{\rm A}(E) \Gamma_{\rm R}(E) ]$, where $\Gamma_{{\rm L/R}}(E)$ is the broadening function defined as $
    \Gamma_{{\rm L/R}}(E) =  {\rm{i}}[ \Sigma_{{\rm L/R}}(E) - \Sigma_{{\rm L/R}}^\dag(E) ]$. The DOS is calculated from following equation:
\begin{equation*}
\textrm{DOS}(E) = \textrm{Im}\left[ G(E) - G^\dag(E) \right] /2\pi. 
\end{equation*}

In this study we only consider coherent transport and ignore internal scattering effects such as inelastic transport or incoherent transport\cite{Nozaki} associated with electron-phonon coupling.\cite{Frederiksen,Ryndyk} 

\section{Analytic solution of surface green function in semi-infinite 1D electrode}
\begin{figure}[ht!]
\begin{center}
\includegraphics[width=8.5cm,clip=true]{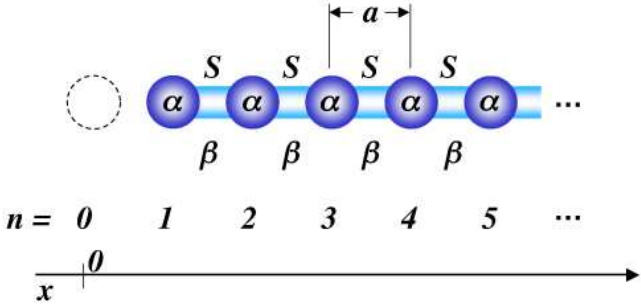}
\end{center}
\caption{\small{Semi-infinite 1D electrode consisting of equally spaced sites with lattice constant $a$. $\alpha$ is on-site energy. $S$ and $\beta$ are hopping integral and overlap matrix between nearest neighbors, respectively. }}
\label{AppB1}
\end{figure}

Here we derive the analytic solution of surface green function of a semi-infinite 1D electrode consisting of equally spaced sites. Consider a semi-infinite 1D electrode with lattice constant $a$ as shown in Fig.~\ref{AppB1}. Each site has only one orbital whose position is given by $x_n=an$ and its on-site energy is $\alpha$. We only consider the interaction between nearest neighbors. The hopping integral and overlap between nearest neighbors are $\beta$ and $S$, respectively. Energy-dependent coupling between nearest neighbors is defined as  $\beta^E=\beta-ES$. 

The wave function of semi-infinite electrode can be constructed from forward- and backward-propagating plane wave $ {\rm{e}} ^{\pm  {\rm{i}} kx}$: 
\begin{equation*}
        \Psi(k) = \frac{1}{\sqrt{2}} ({\rm{e}} ^{ {\rm{i}} kx}- {\rm{e}} ^{- {\rm{i}} kx}).
\end{equation*}
The bra and ket are defined as
\begin{eqnarray*}
     |\Psi(k)\rangle &= \frac{1}{\sqrt{2}} ( {\rm{e}} ^{ {\rm{i}} kx}-{\rm{e}} ^{- {\rm{i}} kx})|x\rangle = \frac{1}{\sqrt{2}} (2  {\rm{i}} \sin kx ) |x\rangle\\
    \langle \Psi(k)| &= \frac{1}{\sqrt{2}} ( {\rm{e}} ^{- {\rm{i}} kx}-{\rm{e}} ^{ {\rm{i}} kx})|x\rangle = \frac{1}{\sqrt{2}} (-2  {\rm{i}} \sin kx ) \langle x|.
\end{eqnarray*}
Thus, the wave function at site $n$ is given by:
\begin{equation*}
\langle x_n |\Psi(k)\rangle =  \frac{1}{\sqrt{2}}\sum_x \langle x_n | \Psi(k) | x \rangle =  \frac{1}{\sqrt{2}}(2 {\rm{i}} \sin kx_n).
\end{equation*}
This function satisfies the boundary condition of semi-infinite 1D electrode that the wave function on site $n=0$ is zero since $ \Psi(x_0)= \langle x_0 | \Psi(k)\rangle=0  $. 

\begin{figure}[ht!]
\begin{center}
\includegraphics[width=6.0cm,clip=true]{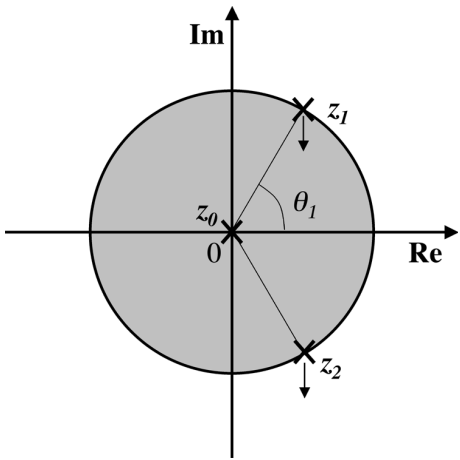}
\end{center}
\caption{\small{Singularity points of the function $f(z)$ on complex variable plane. The gray zone is surrounded by unit circle $z= {\rm{e}} ^{ {\rm{i}} \theta}$. Because of the imaginary term $ {\rm{i}}\delta$ in Eq.~(\ref{eq:oint}), the singularity points, $z_1$ and $z_2$ move to inside and outside of the circle, respectively.}}
\label{DectA2}
\end{figure}

The operator of Green's function is defined as
\begin{equation*}
\hat{G}^{\rm R}(E) = \sum_k \frac{| \Psi(k)\rangle\langle \Psi(k) |}{ E+ {\rm{i}} \delta - \varepsilon(k)}.
\end{equation*}
The free propagator of the electrode between sites $n$ and $n'$ is given by
\begin{equation*}
\label{eq:g0nn}
    (G^{\rm R}(E))_{n,n'} = \sum_k \frac{\langle x_n| \Psi(k)\rangle\langle \Psi(k) |x_{n'}\rangle}{ E+ {\rm{i}}\delta - \varepsilon(k)}.
\end{equation*}
Especially in the case of $n=n'$ we get
\begin{equation*}   
   (G^{\rm R}(E))_{n,n}  =\sum_{k}\frac{2\sin^2 kna}{E+ {\rm{i}} \delta -\varepsilon (k) }. 
\end{equation*}
   We only need the matrix element at the terminal site since this is the only site where the molecular system couples, thus putting $n=1$ gives
\begin{equation*}   
   (G^{\rm R}(E))_{1,1}  =\sum_{k}\frac{2\sin^2 ka}{E+{\rm{i}} \delta -\varepsilon (k) }. 
\end{equation*}   
Hereafter we drop $E$ from $(G^{\rm R}(E))_{1,1}$.
Since the number of sites is very large thus the summation can be converted into integral ($\sum_k\to \frac{a}{2\pi } \int_{-\pi/a}^{\pi/a} {\rm{d}} k$) leading to
\begin{equation*}   
   (G^{\rm R})_{1,1}  =\frac{a}{2\pi} \int_{-\pi/a}^{\pi/a}\frac{2\sin^2 ka}{E+ {\rm{i}} \delta- \varepsilon (k)} {\rm{d}} k.
\end{equation*}
Using E-k relation of 1D electrode consisting of equally spaced sites ($\varepsilon (k) = \alpha +2\beta^E \cos ka $), we get
\begin{equation*}   
    (G^{\rm R})_{1,1} = \frac{a}{\pi} \int_{-\pi/a}^{\pi/a}\frac{\sin^2 ka}{E+ {\rm{i}} \delta- \alpha -2\beta^E \cos ka } {\rm{d}} k.\end{equation*}
When we define $\theta = ka$, the variable of  integral can be changed as
\begin{equation*}   
    (G^{\rm R})_{1,1}  =\frac{1}{\pi} \int_{-\pi}^{\pi}\frac{\sin^2 \theta}{E+ {\rm{i}} \delta- \alpha -2\beta^E \cos \theta}{\rm{d}} \theta.
\end{equation*}
When we define  $z={\rm{e}} ^{ {\rm{i}} \theta}$, then we get  $(z-1/z)^2=-4\sin^2\theta$, $(z+1/z)=2\cos \theta$, and ${\rm{d}} z= {\rm{i}} z {\rm{d}} \theta$. Substituting these expressions into  $(G^{\rm R})_{1,1}$, we obtain
\begin{equation*} 
    \label{eq:int_to_oint}  
    ( G^{\rm R} )_{1,1} = \frac{1}{4 {\rm{i}} \pi \beta^E} \oint \frac{ ( z^2-1)^2   }{ z^2(z^2-2pz+1+ {\rm{i}} \delta) }{\rm{d}} z,
\end{equation*}
where  $p \equiv \frac{E-\alpha}{2\beta^E} = \cos \theta$.
The term  $z^2-2pz+1$ can be changed to $(z-z_1)(z-z_2)$. These $z_1$ and $z_2$ must satisfy following relations: $z_1z_2=1$ and $z_1+z_2=2p$. Note that  $z_1=\frac{1}{z_2}={\rm{e}} ^{ {\rm{i}} \theta_1}$ satisfies these relations.
Then $( G^{\rm R} )_{1,1} $ is written as
\begin{equation}    \label{eq:oint}
  \begin{split}  
   ( G^{\rm R} )_{1,1} &= \frac{1}{4 {\rm{i}} \pi\beta^E} \oint \frac{( z^2-1   )^2 }{ z^2((z-z_1)(z-z_2)+ {\rm{i}}\delta) }{\rm{d}} z\\
  &= \frac{1}{4{\rm{i}} \pi\beta^E} \oint f(z){\rm{d}} z.
    \end{split}  
\end{equation}
We can solve this complex integral using residue theorem:
\begin{equation*} 
    \label{eqbu}  
    ( G^{\rm R} )_{1,1} = \frac{1}{4 {\rm{i}} \pi\beta^E} 2\pi  {\rm{i}} \sum_{k=0}^{n} \textrm{Res}[f(z)]_{z=z_k}.
\end{equation*}
Figure \ref{DectA2} shows the singular points of function $f(z)$ on  complex variable plane. The residual term for Res$[f(z)]_{z=z_2}$ vanishes since the imaginary positive infinitesimal term $ {\rm{i}} \delta$ move the singular point at $z_2$ outside of the circle. Therefore, $( G^{\rm R} )_{1,1}$ is 
\begin{equation*} 
    \label{eqbuv}  
    ( G^{\rm R} )_{1,1} = \frac{1}{2\beta^E} (\textrm{Res}[f(z)]_{z=z_0} + \textrm{Res}[f(z)]_{z=z_1}),
\end{equation*}
where 
\begin{equation*} 
    \label{efa}  
    \textrm{Res}[f(z)]_{z=z_1} = [(z-z_1)f(z)]_{z=z_1}=z_1-z_2,
\end{equation*}
and 
\begin{equation*} 
    \textrm{Res}[f(z)]_{z=z_0} = \left[  \frac{{\rm{d}} }{{\rm{d}} z} \frac{(z^2-1)^2}{(z-z_1)(z-z_2)}  \right]_{z=z_0} = z_1+z_2.
\end{equation*}
Finally, we get the surface green function of 1D electrode consisting of equally spaced sites as
\begin{equation*} 
    \label{efa2}  
    \Sigma_{{\rm L/R}}(E) = V_{{\rm L/R}}^{E}(G_0^{\rm R}(E))_{1,1}V_{{\rm L/R}}^{E\dag}
= V_{{\rm L/R}}^E\frac{ {\rm{e}}^{ {\rm{i}} \theta}}{\beta^E}V_{{\rm L/R}}^{E\dag},
\end{equation*}
where, $V_{{\rm L/R}}^E=V_{{\rm L/R}}-ES_{{\rm L/R}}$ is energy-dependent coupling strength between the electrode and molecule in non-orthogonal description, where $S_{{\rm L/R}}$ is the overlap between the terminal site in the electrode and the molecule.


\begin{thebibliography}{99}
\bibitem{1Gio-Book05} Cuniberti G, Fagas G and Richter K, Eds. 2005 \textit{Introducing Molecular Electronics.} (Springer-Verlag Berlin Heidelberg)
\bibitem{Agrait} Agra\"{i}t N, Levy-Yeyati A and van Ruitenbeek J M 2003 {\it Phys. Rep.} {\bf377} 81 
\bibitem{Hisamoto00} Hisamoto D, Lee W-C, Kedzierski J, Takeuchi H, Asano K, Kuo C, Anderson E,  King T-J, Bokor J and Hu C 2000 {\it IEEE, Trans. Elec. Dev.} {\bf 47} 2320
\bibitem{Pai-Chun06} Chang P-C, Fan Z, Chien C-J, Stichtenoth D, Ronning C and Lu J G 2006 {\it Appl. Phys. Lett.} {\bf 89} 133113
\bibitem{Huang02} Huang Y, Duan X, Cui Y and Lieber C M 2002 {\it Nano Lett.} {\bf 2} 101 
\bibitem{Singh06} Singh N, Agarwal A, Bera L K, Liow T Y, Yang R, Rustagi S C, Tung C H, Kumar R, Lo G Q, Balasubramanian N and Kwong D-L 2006 {\it IEEE, Elec. Dev. Lett.}  {\bf 27} 383
\bibitem{Takayanagi} Ohnishi H, Kondoh Y and Takayanagi K 1998 {\it Nature} {\bf 395}, 780
\bibitem{Yanson} Yanson A I, Bollinger G R, vand den Brom H E, Agra\"{i}t N and van Ruitenbeek  J M 	1998 {\it Nature} {\bf 395} 783
\bibitem{Smit} Smit R H M, Untiedt C, Rubio-Bollinger G, Segers R C and van Ruitenbeek J M 2003 {\it Phys. Rev. Lett.} {\bf 91} 076805
\bibitem{Bettini} Bettini J, Sato F, Coura P Z, Dantas S O, Galv\~{a}o D S and Ugarte D 2006 {\it Nature Nanotechnol.} {\bf 1} 182
\bibitem{Ho-Chain} Nilius N, Wallis T M and Ho W 2003 {\it Phys. Rev. Lett.} {\bf 90} 186102
\bibitem{Cchain} Jin C, Lan H, Peng L, Suenaga K and Iijima S 2009 {\it Phys. Rev. Lett.} {\bf 102} 205501 
\bibitem{Chuvlin} Chuvlin A, Meyer J C, Algara-Siller G and Kaiser U 2009 {\it New J. Phys.} {\bf 11} 083019 
\bibitem{Landauer} Landauer R 1981 {\it Phys. Lett.} {\bf 85A} 91
\bibitem{Buttiker} B\"{u}ttiker M 1986 {\it Phys. Rev. Lett.} {\bf 57} 1761
\bibitem{Ellenbogen} Ellenbogen J C and Love J C 2000 {\it Proceedings of the IEEE} {\bf 88} 386
\bibitem{Joachim} Joachim C, Gimzewski J K and Aviram A 2000 {\it Nature} {\bf 408} 541
\bibitem{Cui01} Cui Y, Wei Q, Park H and Lieber C M 2001 {\it Science} {\bf 293} 1289
\bibitem{Patolsky04} Patolsky F, Zheng G, Hayden O, Lakadamyali M, Zhuang X and Lieber C M 2004 {\it Proc. Natl. Acad. Sci. USA} {\bf 101} 14017
\bibitem{Moleuclar-wire} De Cola L Ed. \textit{Molecular Wires: From Design to Properties. } {\it Top. Curr. Chem.} {\bf 257} (Springer-Verlag Berlin Heidelberg, 2005).
\bibitem{Israel-STM} Calev Y, Cohen H, Cuniberti G, Nitzan A and Porath D 2004 {\it Israel J. Chem.} {\bf 44} 133
\bibitem{NDLang} Lang N D 1997 {\it Phys. Rev. Lett.} {\bf 79} 1357
\bibitem{Ubiquitous} Forrest S R 2004 {\it Nature} {\bf 428} 911
\bibitem{PPV} Burroughes J H, Bradley D D C, Brown A R, Marks R N, McKay K, Friend R H, Burns P L and Holmes A 1990 {\it Nature} {\bf 347} 539 
\bibitem{OLED} Friend R H, Gymer R W, Holmes A B, Burroughes J H, Marks R N, Taliani C, Bradley D D C, Dos Santos D A, Br\'{e}das J L, L\"{o}gdlund M and Salaneck W R 1999 {\it Nature} {\bf 397} 121
\bibitem{Jenekhe} Kulkarni A P, Tonzola C J, Babel A and Jenekhe S A 2004  {\it Chem. Mater.}  {\bf 16} 4556  
\bibitem{TFT} Katz H E, Bao Z and Gilat S L 2001 {\it Acc. Chem. Res.} {\bf 34} 359
\bibitem{Graetzel} Gr\"{a}tzel M 2001 {\it Nature} {\bf 414} 338
\bibitem{Gunes} G\"{u}nes S, Neugebauer H and Sariciftci N S 2007  {\it Chem. Rev.} {\bf 107} 1324 
\bibitem{Blom} Blom P W M, Mihailetchi V D, Koster L J A and Markov D E 2007 {\it Adv. Mater.} {\bf 19} 1551  
\bibitem{Chiang77} Chiang C K, Fincher Jr. C R, Park Y W, Heeger A J, Shirakawa H, Louis E-J, Gau S C and MacDiarmid A G 1977 {\it Phys. Rev. Lett.} {\bf 39} 1098 
\bibitem{Chiang78} Chiang C K, Dury M A, Gau S C, Heeger A J, Louis E-J, MacDiarmid A G, Park Y W and Shirakawa H 1978 {\it J. Am. Chem. Soc.} {\bf 100} 1013  
\bibitem{Brian} Gregg B A 2009 {\it J. Phys. Chem. C} {\bf 113} 5899 
\bibitem{SSH79} Su W P, Schrieffer J R and Heeger A J 1979 {\it Phys. Rev. Lett.} {\bf 42} 1698
\bibitem{Mujica1} Mujica V, Kemp M and  Ratner M A 1994 {\it J. Chem. Phys.}  {\bf 101}  6849 
\bibitem{Mujica2} Mujica V, Kemp M and Ratner M A 1994 {\it J. Chem. Phys.} {\bf 101}  6856 
\bibitem{PAW95} Pastawski H M, Weisz J F and Albanesi E A 1995 {\it Phys. Rev. B} {\bf 52} 10665
\bibitem{PM01} Pastawski H M and Medina E 2001 {\it Rev. Mex. Fis.} {\bf 47s1} 1
\bibitem{Tyler} Tyler D R 2007 {\it Frontiers in transition Metal-Containing Polymers} (Wiley-Interscience, Weinheim, Germany)
\bibitem{Bera} Bera J K and Dunbar K R 2002 {\it Angew. Chem. Int. Ed.} {\bf 41} 4453 
\bibitem{Peierls} Peierls R E 1955 {\it Quantum Theory of Solids} (Clarendon Press; Oxford, U.K.) 
\bibitem{Conwell} Conwell E M 1988 {\it Semiconductors and Semimetals} (Academic Press: Dordrecht, Netherlands)
\bibitem{Anderson1} Anderson P W 1961 {\it Phys. Rev.} {\bf124} 41
\bibitem{Anderson2} Newns D M 1969  {\it Phys. Rev.} {\bf 178} 1123
\bibitem{Azbel} Ricco B and Azbel M Y 1984 {\it Phys. Rev. B} {\bf 29} 1970
\bibitem{Luis} Fo\`{a} Torres L E F, Pastawski H M and Makler S S 2001 {\it Phys. Rev. B} {\bf 29} 193304
\bibitem{DPW89} D'Amato J L, Pastawski H M and Weisz J F 1989 {\it Phys. Rev. B} {\bf 39} 3554
 \bibitem{LPD90}  Levstein P R, Pastawski H M and D'Amato J L 1990 {J. Phys.: Condens. Matter} {\bf 2} 1781

\bibitem{Emberly} Emberly E G and Kirczenow G 1998 {\it Phys. Rev. B} {\bf 58} 10911 
\bibitem{Fisher-Lee} Fisher D S and Lee P A 1981 {\it Phys. Rev. B} {\bf 23} 6851
\bibitem{Nozaki} Nozaki D, Girard Y and Yoshizawa K 2008 {\it J. Phys. Chem. C} {\bf 112} 17408
\bibitem{Ryndyk} Ryndyk D A and Cuniberti G 2007 {\it Phys. Rev. B} {\bf 76} 155430 
\bibitem{Frederiksen} Frederiksen T, Brandbyge M and Jauho A-P 2004 {\it J. Comp. Elec.} {\bf 3} 423



\end{thebibliography}
\end{document}